\documentclass[twocolumn,prl,aps,showpacs]{revtex4}
\usepackage{graphicx}

\begin{document}

\newcommand{\snn}{\sqrt{s_{NN}}}
\newcommand{\seff}{\sqrt{s_{\rm eff}}}
\newcommand{\s}{\sqrt{s}}
\newcommand{\pp}{pp}
\newcommand{\pbarp}{\overline{p}p}
\newcommand{\qbarq}{\overline{q}q}
\newcommand{\epem}{e^+e^-}

\newcommand{\nhit}{N_{hit}}
\newcommand{\npp}{n_{pp}}
\newcommand{\nch}{N_{ch}}
\newcommand{\np}{N_{part}}
\newcommand{\ns}{N_{spec}}
\newcommand{\ntot}{\langle\nch\rangle}
\newcommand{\avenp}{\langle\np\rangle}
\newcommand{\npB}{N_{part}^B}
\newcommand{\nc}{N_{coll}}
\newcommand{\avenc}{\langle\nc\rangle}
\newcommand{\half}{\frac{1}{2}}
\newcommand{\halfnp}{\langle\np/2\rangle}
\newcommand{\etap}{\eta^{\prime}}
\newcommand{\as}{\alpha_{s}(s)}
\newcommand{\etazero}{\eta = 0}
\newcommand{\etaone}{|\eta| < 1}
\newcommand{\dndeta}{d\nch/d\eta}
\newcommand{\dndetazero}{\dndeta|_{\etazero}}
\newcommand{\dndetaone}{\dndeta|_{\etaone}}
\newcommand{\dndetanp}{\dndeta / \halfnp}
\newcommand{\dndetaonp}{\dndeta / \np}
\newcommand{\dndetazeronp}{\dndetazero / \halfnp}
\newcommand{\dndetaonenp}{\dndetaone / \halfnp}
\newcommand{\ratio}{\ntot/\halfnp}
\newcommand{\nee}{N_{\epem}}
\newcommand{\nhh}{N_{hh}}
\newcommand{\nubar}{\overline{\nu}}

\title{Comparison of the Total Charged-Particle Multiplicity in High-Energy Heavy Ion Collisions with $\epem$ and $\pp/\pbarp$ Data}

\author{
% Authors for data EXCLUSIVELY from RUN2001-2 which included AuAu @ 20, 200 GeV
% and pp @ 200 GeV
% Note that there is a separate list for papers including 130 GeV AuAu as well
%
%Last edited 25-Apr-2002 by George Stephans\\  \vspace{0.2in}
%
B.B.Back$^1$,
M.D.Baker$^2$,
D.S.Barton$^2$,
R.R.Betts$^6$,
M.Ballintijn$^4$,
A.A.Bickley$^7$,
R.Bindel$^7$,
A.Budzanowski$^3$,
W.Busza$^4$,
A.Carroll$^2$,
M.P.Decowski$^4$,
E.Garc\'{i}a$^6$,
N.George$^{1,2}$,
K.Gulbrandsen$^4$,
S.Gushue$^2$,
C.Halliwell$^6$,
J.Hamblen$^8$,
G.A.Heintzelman$^2$,
C.Henderson$^4$,
D.J.Hofman$^6$,
R.S.Hollis$^6$,
R.Ho\l y\'{n}ski$^3$,
B.Holzman$^2$,
A.Iordanova$^6$,
E.Johnson$^8$,
J.L.Kane$^4$,
J.Katzy$^{4,6}$,
N.Khan$^8$,
W.Kucewicz$^6$,
P.Kulinich$^4$,
C.M.Kuo$^5$,
W.T.Lin$^5$,
S.Manly$^8$,
D.McLeod$^6$,
J.Micha\l owski$^3$,
A.C.Mignerey$^7$,
R.Nouicer$^6$,
A.Olszewski$^3$,
R.Pak$^2$,
I.C.Park$^8$,
H.Pernegger$^4$,
C.Reed$^4$,
L.P.Remsberg$^2$,
M.Reuter$^6$,
C.Roland$^4$,
G.Roland$^4$,
L.Rosenberg$^4$,
J.Sagerer$^6$,
P.Sarin$^4$,
P.Sawicki$^3$,
W.Skulski$^8$,
S.G.Steadman$^4$,
P.Steinberg$^2$,
G.S.F.Stephans$^4$,
M.Stodulski$^3$,
A.Sukhanov$^2$,
J.-L.Tang$^5$,
R.Teng$^8$,
A.Trzupek$^3$,
C.Vale$^4$,
G.J.van~Nieuwenhuizen$^4$,
R.Verdier$^4$,
B.Wadsworth$^4$,
F.L.H.Wolfs$^8$,
B.Wosiek$^3$,
K.Wo\'{z}niak$^3$,
A.H.Wuosmaa$^1$,
B.Wys\l ouch$^4$\\
\vspace{3mm}
\small
%
% Note that this is the full form of the addresses, for conference proceedings,
% you can use the reduced one that follows
%
% $^1$~Physics Division, Argonne National Laboratory, Argonne, IL 60439-4843,
% USA\\
% $^2$~Chemistry and C-A Departments, Brookhaven National Laboratory, Upton, NY
% 11973-5000, USA\\
% $^3$~Institute of Nuclear Physics, Krak\'{o}w, Poland\\
% $^4$~Laboratory for Nuclear Science, Massachusetts Institute of Technology,
% Cambridge, MA 02139-4307, USA\\
% $^5$~Department of Physics, National Central University, Chung-Li, Taiwan\\
% $^6$~Department of Physics, University of Illinois at Chicago, Chicago, IL
% 60607-7059, USA\\
% $^7$~Department of Chemistry, University of Maryland, College Park, MD 20742,
% USA\\
% $^8$~Department of Physics and Astronomy, University of Rochester, Rochester,
% NY 14627, USA\\
%
%
$^1$~Argonne National Laboratory, Argonne, IL 60439-4843, USA\\
$^2$~Brookhaven National Laboratory, Upton, NY 11973-5000, USA\\
$^3$~Institute of Nuclear Physics, Krak\'{o}w, Poland\\
$^4$~Massachusetts Institute of Technology, Cambridge, MA 02139-4307, USA\\
$^5$~National Central University, Chung-Li, Taiwan\\
$^6$~University of Illinois at Chicago, Chicago, IL 60607-7059, USA\\
$^7$~University of Maryland, College Park, MD 20742, USA\\
$^8$~University of Rochester, Rochester, NY 14627, USA\\
}
\date{\today}

\begin{abstract}

The PHOBOS experiment at RHIC has measured the total multiplicity of
primary charged particles as a function of collision centrality
in Au+Au collisions at $\snn = $ 19.6, 130 and 200 GeV.
Above $\s\approx 20$ GeV, the total multiplicity per participating
nucleon pair ($\ratio$) in central events scales with $\s$
in the same way as $\ntot$ in $\epem$ data.
This is suggestive of a universal mechanism of particle production 
in strongly-interacting systems, controlled mainly by the amount of energy
available for particle production 
(per participant pair for heavy ion collisions).
The same effect has been observed in $\pp/\pbarp$ data after correcting
for the energy taken away by leading particles.
An approximate independence of $\ratio$ on the number of participating
nucleons is also observed, reminiscent of ``wounded nucleon''
scaling ($\nch \propto \np$), 
but with the constant of proportionality set by the 
multiplicity measured in $\epem$ data rather than by $\pp/\pbarp$ data.

\end{abstract}
\maketitle

Central collisions of two gold nuclei
at the top energy of the Relativistic Heavy Ion Collider (RHIC) at
Brookhaven National Laboratory produce thousands of charged particles.
These are the largest particle multiplicities generated
in man-made subatomic reactions.  The hope is that these
complex systems may reveal evidence of the creation and decay of 
a Quark-Gluon Plasma (QGP), 
where quarks and gluons are allowed to explore a volume larger than that of
a typical hadron.

The high multiplicities in heavy ion collisions typically
arise from the large number of nucleon-nucleon collisions
which occur, with many of the nucleons struck several times 
as they pass through the oncoming nucleus. 
Studies of proton-nucleus collisions demonstrated that the
total multiplicity ($\nch$) does not scale proportionally to
the number of binary collisions ($\nc$) in the reaction,
but rather was found to scale more closely with the number of
``wounded nucleons'' which participate inelastically ($\np$)
\cite{elias,wounded}.
For example, the number of participants is $\np=2$ for a 
proton-proton collision and $\np=(\nc+1)$ for a proton-nucleus collision.
Thus,  by scaling the particle yields measured in heavy ion collisions
by $\np/2$, data from heavy ion collisions may be directly compared with
similar yields in elementary $\pp$, $\pbarp$ or even 
the annihilation of $\epem$ into hadrons.  

While both $\epem$ and $\pp/\pbarp$ collisions 
must ultimately allow a description based
on Quantum Chromodynamics (QCD), the theory of the strong
interaction, the evolution of these two systems tends to be
understood in different ways.
The large momentum transfer to the outgoing produced quark
and anti-quark in $\epem$ reactions
allows the use of perturbative QCD (pQCD) to describe the 
spectrum of quarks and gluons radiated as the system fragments
\cite{basics}.
Minimum bias collisions of hadrons 
are not generally thought to be amenable
to such a perturbative description, since the transverse
momentum exchanges involved are typically less than 1 GeV/c.
Instead, phenomenological approaches (e.g. PYTHIA \cite{pythia}) 
are used to describe most of the (predominantly soft) particles 
produced in high energy $\pp$ or $\pbarp$ collisions.

A basic connection between perturbative and non-perturbative
physics has been elucidated 
by simultaneous measurements of the multiplicity
of charged particles and the high-momentum ``leading'' protons
in $\pp$ collisions at the ISR.
Basile {\it et al.} \cite{basile} 
found that the average multiplicity $\ntot$ in $\pp$
collisions is similar to that for $\epem$ collisions with
$\s_{\epem} = \seff$, where $\seff$ is the $\pp$ center-of-mass 
energy minus the energy of the leading particles.
This is interpreted as a universal
mechanism of particle production controlled dominantly by 
the available center of mass energy \cite{basile}.

In this Letter, we report results from the PHOBOS experiment on the 
total multiplicity of primary charged particles $\ntot$ as a function of $\np$
for heavy ion collisions at $\snn=$ 19.6, 130 and 200 GeV, where
$\snn$ is the nucleon-nucleon center-of-mass energy.
Comparisons with $\pp/\pbarp$ and $\epem$ data are made
to investigate whether this universal mechanism of
particle production applies in the context of heavy ion collisions.

The PHOBOS multiplicity detector consists of two arrays of silicon
detectors which cover nearly the full solid angle for collision
events.
The ``Octagon'' detector surrounds the interaction region
with a roughly cylindrical geometry covering $|\eta|<3.2$.  
Two sets of three ``Ring'' detectors are placed far forward and backward
of the interaction point and surround the beam pipe, covering $3<|\eta|<5.4$.
The methods used for measuring the multiplicity of charged particles 
as well as for extracting $\avenp$ 
have been described in more detail in 
Refs. \cite{phobos_cent_200,phobos_limfrag}.

\begin{figure}
\begin{center}
\includegraphics[width=9cm]{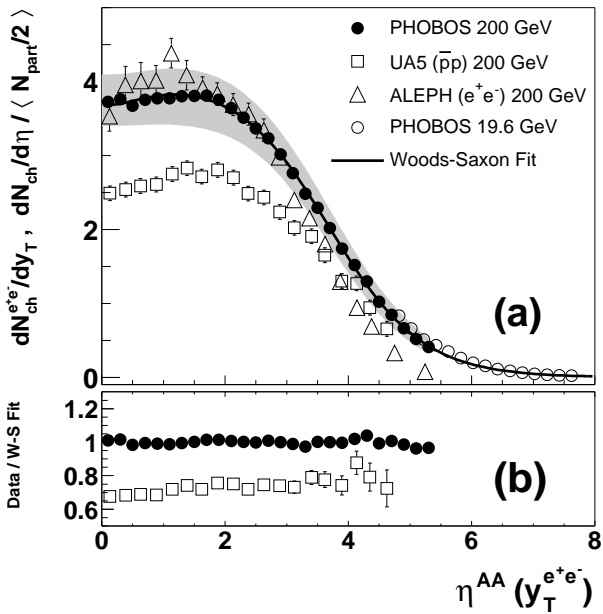}
\end{center}
\caption{
(a) $\dndetanp$ of charged particles produced
in central Au+Au collisions at $\snn=$ 200 and 19.6 GeV
(shifted by $\Delta\eta=2.32$), compared with elementary
systems.  A Woods-Saxon fit to the 200 GeV Au+Au data is shown.
The $\epem$ data are plotted as a function of $y_T$, the rapidity
relative to the thrust axis.
(b) PHOBOS and UA5 data divided by a
Woods-Saxon fit to the 200 GeV Au+Au data.
}
\label{figure1}
\end{figure}

Using the data presented in Ref. \cite{phobos_limfrag},
Fig. \ref{figure1}a shows 
$\dndeta / \avenp/2$ averaged over the forward and
backward hemispheres for the most central 
Au+Au events 
at $\snn=$ 200 GeV. 
The systematic errors (representing a 90\% CL interval) 
depend on $\eta$ and are shown on the figure as a shaded band.
The Au+Au data are compared with $\dndeta$ for 
non-single diffractive (NSD) $\pbarp$ collisions \cite{ua5} and 
$dN/dy_T$ for $\epem$ collisions (with cuts applied to
reject large initial-state photon radiation) \cite{ALEPH} 
at $\s = $ 200 GeV.
The variable $y_T$ is the rapidity of charged particles 
relative to the  event thrust axis, assuming all particles to
have the pion mass.

It is observed that the Au+Au data are very similar in
magnitude and shape to the $\epem$ data at the same
$\s$, and similar in shape to the $\pbarp$ data (as shown 
in Fig. \ref{figure1}b), over a large range in $\eta$.
The differences between the $\epem$ and Au+Au distributions
shown in Fig. \ref{figure1}a can be partly attributed to
the different kinematic variables.
JETSET calculations indicate that the $y_T$ distribution
is slightly narrower than the corresponding
pseudorapidity distribution in $\epem$ collisions, with
a higher plateau height.  Yet, even without taking this into account,
the difference between the distributions is no more than
$\pm10\%$ for $|\eta|$ and $|y_T|<4$ \cite{pythia}.
However, the same calculations also show that the 
choice in kinematic variables does not explain the 
difference in the forward region (above $|\eta|=4$),
although this may not be surprising, as this region 
should show some residual effect of the presence of the
spectator nucleons.

The similarity of the angular distributions indicates that the total
yield of charged particles in $\epem$ and central Au+Au collisions 
should also be similar for the same $\s$, when the nuclear data 
are scaled by the number of participant pairs.  
To correct for the small acceptance losses in the PHOBOS
apparatus (which covers $|\eta|<5.4$),
we have used several methods inspired by the excellent agreement of
the lowest energy PHOBOS data with the higher energy data
when shown as a function of  $\etap = \eta-y_{beam}$ \cite{phobos_limfrag}.
PHOBOS data from $\snn=19.6$ GeV for $\eta>2.5$, 
shifted by $\Delta\eta=y_{200}-y_{19.6}=2.32$
(the difference in beam rapidities between the two energies),
displays the limiting behavior discussed in Ref. \cite{phobos_limfrag}.
This effectively extends the rapidity coverage to $\eta \sim 8$.
A Woods-Saxon function for $dN/dy$ fit to the Au+Au data,
also provides a reasonable description of the $dN/d\eta$ distribution,
and extrapolates through the lower energy data as well.
Thus, in one method, we integrate $\dndeta$ for $\snn = $ 130 and 200
GeV for $\etap<0$ and then use the
PHOBOS data at $\snn$ = 19.6 GeV for $\etap>0$.
We also integrate Woods-Saxon fits, similar to that shown in
Fig. \ref{figure1}a, for $|\eta|<8$.
These two approaches agree within 2\% for central events.
For the lowest RHIC energy, we simply integrate the charged particles
in the PHOBOS acceptance.

\begin{figure}
\begin{center}
\includegraphics[width=9cm]{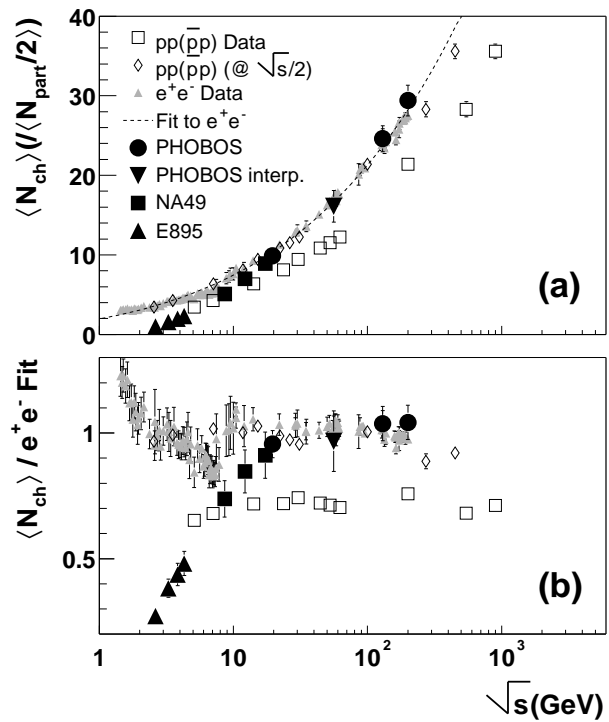}
\end{center}
\caption{
(a) The total charged multiplicity $\ntot$ for $\pp$, $\pbarp$,
$\epem$ and central Au+Au events as a function of $\s$.  
The Au+Au data are normalized by $\np/2$.
The dotted line is a perturbative QCD expression fit to the $\epem$ data.  
The diamonds are the $\pp$/$\pbarp$ data with $\seff=\s/2$.
(b) The data in (a) divided by the $\epem$ fit, to allow direct
comparison of different data at the same $\s$.
}
\label{figure2}
\end{figure}

In Fig. \ref{figure2}a, data on $\nch$ from $\pp$, $\pbarp$, $\epem$
and central heavy ion collisions (scaled by $\np/2$)
are shown as a function of $\s$.  
The $\pp$, $\pbarp$, and $\epem$ data and errors are taken from a compilation 
\cite{Groom:in} and no further corrections are applied.
The errors shown are the quadratically combined statistical and systematic
errors.
Heavy ion data are shown for central Au+Au events at RHIC 
(this work), 
Au+Au events from E895 at the AGS ($\snn=2.6-4.3$ GeV) \cite{e895} and
Pb+Pb events from NA49 at the SPS ($\snn=$ 8.6, 12.2 and 17.3 GeV) \cite{na49}.
A PHOBOS Au+Au data point at $\snn=56$ GeV has been added by using 
the measured value at midrapidity \cite{phobos_dndeta} and 
using the universal limiting distribution described in 
Ref. \cite{phobos_limfrag} to approximate the shape of the full distribution.
All of the errors shown for the heavy ion data are systematic.
 
Perturbative QCD calculations
are able to predict the
dependence of the total multiplicity in $\epem$ collisions as a function of
$\s$, $N_{\epem}(s) = C \as^A\exp(\sqrt{B/\as})$,
with $A$ and $B$ fully calculable within pQCD\cite{Mueller:cq}. 
The QCD scale $\Lambda_{QCD}$ is set to 225 MeV, leaving
only a constant of proportionality $C$ free to fit to
the experimental data.
A fit to the $\epem$ data has been made with
this expression (``$\epem$ fit'') and has been
used in Fig. \ref{figure2}b to see how the various systems
compare with $\epem$ data 
by scaling all of the data at a given $\s$ by this function.

Fig. \ref{figure2}b shows that the $\pp/\pbarp$ data
are about 30\% below $\epem$ over the full range of energies.  
However, rescaling the $\s$ of each point by a factor of $1/2$,
$\seff=\s/2$,
brings the data into reasonable agreement with the $\epem$ trend,
as shown by the open diamonds.
This is consistent with measurements of leading protons 
in $\pp$ collisions, which find $dN/dx_F$ (where $x_F$ = $2p_z/\s$
in the collider reference frame) 
to be approximately constant
for non-diffractive events over a large range of $\s$
\cite{batista} and thus $\langle x_F \rangle\sim 1/2$.

Unlike the $\pp/\pbarp$ data, the heavy ion data does not
follow the $\epem$ trend over the whole energy range.
Instead, they lie below the $\pp$ data at AGS energies, 
crosses through the $\pp$ data between AGS and SPS energies,
and joins smoothly with the $\epem$ data above the top SPS energy.
Thus, at high energies, the multiplicity measured per participant
pair in Au+Au collisions scales in a similar way to $\epem$ data 
at the {\it same} $\s$.
If we understand the lower effective $\s$ in $\pp$ collisions
as stemming from the ``leading particle effect'', where the leading
proton carries off a substantial amount of the available energy,
the Au+Au data suggest a substantially reduced leading
particle effect 
in central collisions of heavy nuclei at high energy.

The alleviation of the leading particle effect 
might not be so surprising in central nuclear collisions.
In the Glauber model, each participating nucleon is typically
struck $4-6$ 
times on average as it passes through the oncoming gold nucleus
in a central event (the exact value of $\nubar$ depending on the 
energy-dependent nucleon-nucleon inelastic cross section, $\sigma_{NN}(s)$).
One could speculate that the multiple
collisions simultaneously 
excite and dissociate the participating nucleons, 
transferring much more of the energy from
the forward direction towards midrapidity than found in
an average $\pp/\pbarp$ collision - but ultimately
limited by the total incident energy.
This hypothesis should be testable in proton-nucleus collisions,
by measuring particle yields as a function of $\nu$ as was done
in Refs. \cite{e910,na49-pp}.
The data in those references suggest that pion 
yields, whether in the projectile region ($y>0$) \cite{na49-pp}
or integrated over $4\pi$ \cite{e910} increase rapidly for 
$\nu<3$ and then much more slowly for $\nu > 3$.
However, limited 
experimental acceptances and theoretical uncertainties preclude
making any strong conclusions regarding the relationship between
the energy loss of the projectile and the total charged multiplicity.

\begin{figure}
\begin{center}
\includegraphics[width=8.5cm]{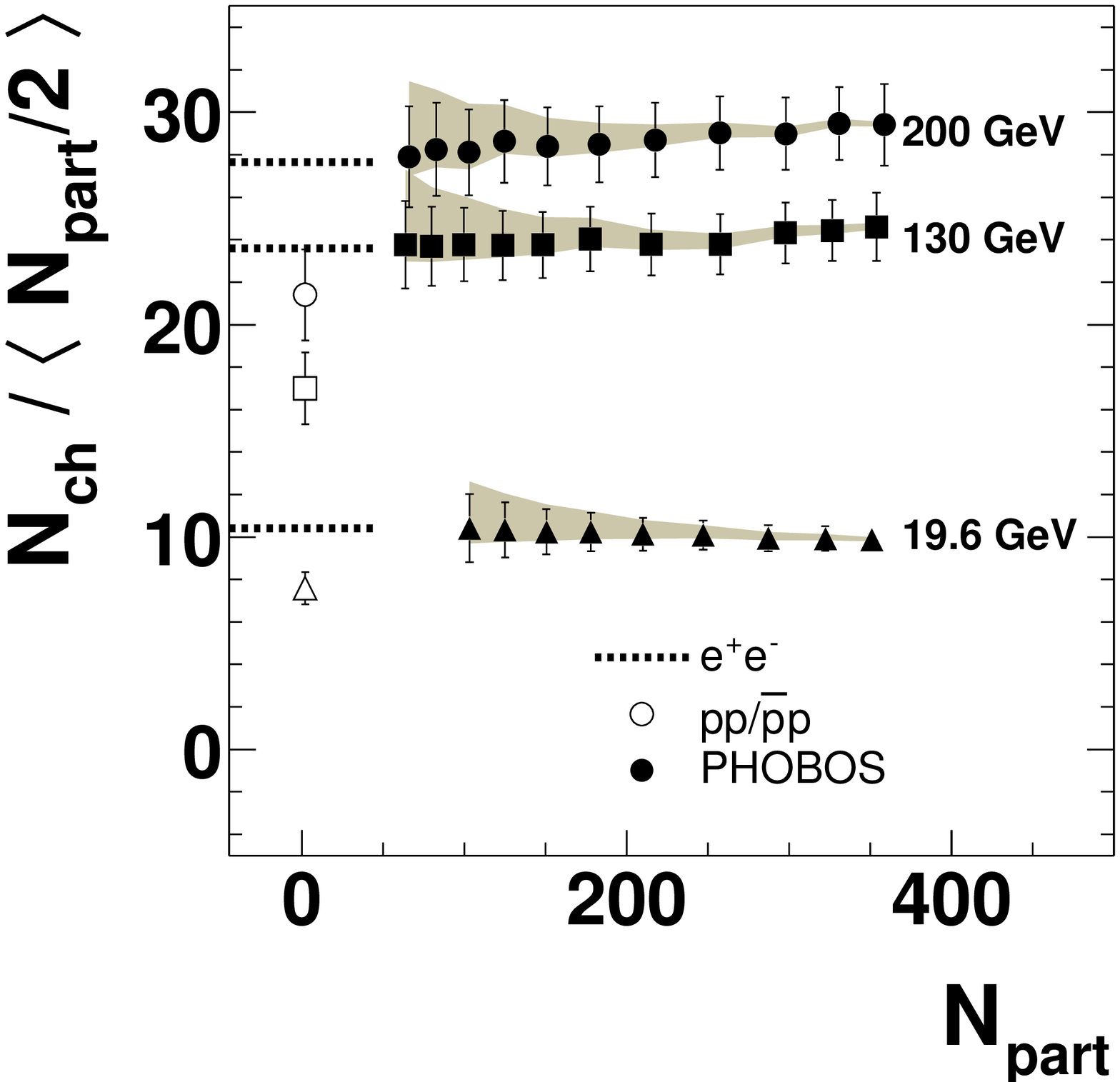}
\end{center}
\caption{
$\ratio$ is shown vs. $\np$ for $\snn = $ 19.6, 130, and 200 GeV
as closed symbols.  The error includes contributions from the
uncertainty on overall $\nch$ scale and $\np$ scale.  The shaded
band shows the uncertainty on the extrapolation procedure.
The open symbols show UA5 data at 200 GeV
and results from an interpolation for the lower energies.
The dotted lines show the values from the $\epem$ fit.
}
\label{figure3}
\end{figure}

In Fig. \ref{figure3} $\ratio$ is shown
for PHOBOS data at three RHIC energies as a function of $\np$.
The 90\% CL systematic error on the centrality dependence of $\ratio$ 
is shown as a shaded band, and represents
a combination of several factors, dominated by
the uncertainty of the extrapolation procedure to extract $\nch$
over the full solid angle.

It might have been expected that, in events
with larger impact parameters, each participant would have
fewer collisions on average and thus not be fully dissociated.
However, within the systematic errors, the total yield per participant pair is 
approximately constant (within 10\%) over the measured centrality range,
$65 < \avenp < 358$, which corresponds to $3 < \nubar < 6$, where
$\nubar$ is the average number of collisions undergone
by each oncoming nucleon.
Thus, it appears that only the first few collisions have any appreciable
effect on particle production.
It should be noted that this simple scaling is not observed for 
particle yields measured in a limited pseudorapidity range near
midrapidity \cite{phobos_cent_200}.

Proton-antiproton data exist at 200 GeV, but not for the other
two RHIC energies.  We use a parametrization of $\pp$ data from Ref. 
\cite{heiselberg}, $\ntot=-4.2+4.69s^{.155}$, for 19.6 and 130 GeV.  
Several measurements  exist in $\epem$ at 200 GeV, but not for
the other two energies.  For these we use the pQCD formula for $\nee$, 
the quality of the fit clearly indicated 
in Fig. \ref{figure2}b.
Fig. \ref{figure3} shows that the Au+Au data are consistent
with ``wounded nucleon'' scaling, 
in that the multiplicity is proportional to $\np$ ($\nch \propto \np$).
However, $\nch$ clearly
does {\it not} scale simply with the multiplicity measured in $\pp$ 
collisions at the same energy.  Rather, for a large range of 
impact parameter, the multiplicity scales approximately with
the total multiplicity in $\epem$ annihilation at the same $\s$.
Thus, it appears that the first few collisions per participant are
sufficient to liberate as much energy for particle production 
as an $\epem$ reaction.

However, the rapid approach of $\ratio$ in 
central heavy ion collisions below $\snn\sim 20$ GeV
towards the $\epem$ data 
complicates any simple geometric interpretation, as all of
the heavy ion data compared are for a similar range of impact parameters.  
One feature that might point to why the particle yields at the
AGS and SPS are perhaps ``suppressed'' relative to $\epem$ data
(and even to $\pp$ data at lower energies, as noted in Ref. \cite{na49})
is the magnitude of the ratio of net baryons to pions in the system.
This ratio, which scales approximately as $\np/\nch$,
is $O(50\%)$ at AGS energies \cite{e895}, but is $O(1\%)$ at RHIC
\cite{phenix_lambda}.  In a thermal statistical 
approach \cite{Cleymans:2002mp}, 
this reflects the decrease of the baryon chemical potential
with increasing beam energy.

In conclusion, the PHOBOS experiment has measured the
normalized charged-particle multiplicity
$\ratio$ in Au+Au collisions at three RHIC energies
as a function of the centrality of the collision.
Above CERN SPS energies, the total multiplicity per participating
nucleon pair, $\ratio$, in central events scales with $\s$
in the same way as $\epem$ data.
This is suggestive of a universal mechanism of particle production 
in strongly-interacting systems, controlled mainly by the amount of energy
available for particle production.
This may be related to the multiple collisions
suffered by each participant nucleon, which could substantially reduce
the leading particle effect seen previously in $\pp$ collisions.
The weak centrality dependence for $\ratio$, reminiscent of 
``wounded nucleon'' scaling,
suggests that after the first few collisions per participant, 
the multiplicity per participant pair saturates near the value measured 
in $\epem$ reactions.
Ultimately, the existence of simple scaling behavior with $\seff$ and $\np$
indicates stronger constraints on particle production than
previously considered theoretically. 
Thus, these results may provide a new perspective on 
particle production in heavy ion collisions.

This work was partially supported by U.S. DOE grants DE-AC02-98CH10886,
DE-FG02-93ER40802, DE-FC02-94ER40818, DE-FG02-94ER40865, DE-FG02-99ER41099, and
W-31-109-ENG-38 as well as NSF grants 9603486, 9722606 and 0072204.  The Polish
groups were partially supported by KBN grant 2 PO3B 103 23.  The NCU group was
partially supported by NSC of Taiwan under contract NSC 89-2112-M-008-024.

\end{document}